\documentclass{mem}
\usepackage{natbib}
\usepackage{txfonts}
\usepackage{balance}
\usepackage{graphicx}
\usepackage{flushend}
\usepackage{xcolor}
\usepackage[breaklinks,pdftex]{hyperref}
\idline{75}{282}
\begin{document}
\def\msun{$\rm M_{\odot}$}
\def\teff{$T\rm_{eff }$}
\def\kms{$\mathrm {km s}^{-1}$}
\newcommand{\stef}{\textcolor{blue}}
\newcommand{\asa}{\textcolor{teal}}
\newcommand{\asacom}[1]{\color{teal}[ÁS: #1]\color{black}}
\newcommand{\andre}{\textcolor{purple}}

\title{
First stars signatures in high-z absorbers
}

   \subtitle{}

\author{
S. \,Salvadori\inst{1,2}, V. D'Odorico\inst{3,4,5},  A. Saccardi\inst{6}, Á. \,Skúladóttir\inst{1,2}, \and I. \,Vanni\inst{1, 2}}

\institute{
Dipartimento di Fisica e Astronomia, Università degli Studi di Firenze, Via G. Sansone 1, 50019, Sesto Fiorentino, Italy
\and
INAF - Osservatorio Astrofisico di Arcetri, Largo E.Fermi 5, 50125, Firenze, Italy
\and
Scuola Normale Superiore, Piazza dei Cavalieri 7, 56126 Pisa, Italy
\and
INAF - Osservatorio Astronomico di Trieste, via G.B. Tiepolo, 11 34143 Trieste, Italy
\and
Institute for Fundamental Physics of the Universe, Via Beirut 2, 34014 Trieste, Italy
\and
{GEPI, Observatoire de Paris, Université PSL, CNRS, 5 Place Jules Janssen, 92190 Meudon, France} \\
\email{stefania.salvadori@unifi.it}
}

\authorrunning{Salvadori S.}

\titlerunning{First stars signatures in high-z absorbers}

\date{Received: Day Month Year; Accepted: Day Month Year}

\abstract{The first stars were likely more massive than those forming today and thus rapidly evolved, exploding as supernovae and enriching the surrounding gas with their chemical products. In the Local Group, the chemical signature of the first stars has been identified in the so-called Carbon-Enhanced Metal-Poor stars (CEMP-no). On the contrary, a similar C-excess was not found in dense neutral gas traced by high-redshift absorption systems. Here we discuss the recent discovery of three C-enhanced very metal-poor ([Fe/H]$<-2$) optically thick absorbers at redshift $z\approx 3-4$, reported by \citep{Saccardi23}.. We show that these absorbers are extra-galactic tracers of the chemical signatures of the first stars, analogous to the CEMP-no stars observed in the Galactic halo and ultra-faint dwarf galaxies. Furthermore, by comparing observations with model predictions we demonstrate that these systems have most likely been imprinted by first stars exploding as low-energy supernovae, which provided $\geq 50\%$ of the metals in these absorbers.
}
\maketitle{}

\section{Introduction}
The first (Pop~III) stars are predicted to be more massive than present-day stars, with masses possibly ranging from some tenth up to thousand times solar \citep[e.g.][]{Hirano2014,Susa2014} and a characteristic mass $>1$\,\msun \citep{Rossi2021}. Massive first stars exploded as supernovae (SNe) and thus contaminated the surrounding environment with their chemical elements, whose yields vary according to the mass of the progenitor star and the SN explosion energy \cite[e.g.][]{Heger2010}. Low-mass stars, born from this metal-polluted gas, can survive until today \citep[e.g.][]{Bromm2001} and retain in their atmospheres a record of the chemical abundances left by the first stars. 

These chemical signatures have been looked for in ancient metal-poor stars in our cosmic neighbourhood: mostly the Milky Way (MW) halo and Local Group dwarf galaxies, where we can uniquely study individual stars \cite[e.g.][]{Tolstoy2009}. 
Very metal-poor stars ($\rm[Fe/H]<-2$)\footnote{\rm [Fe/H]=$\log{\left(\tt \frac{N_{Fe}}{N_H}\right)}-\log{\left(\tt \frac{N_{Fe}}{N_H}\right)_{\odot}}$}, strongly enhanced in carbon over iron ($\rm[C/Fe]>+0.7$) and not enriched in neutron-capture elements ($\rm[Ba/Fe]<0$), i.e.~the so-called ``CEMP-no" stars \citep[e.g.][]{Beers2005}, are the most promising descendants of Pop~III stars that exploded as primordial low-energy supernovae (e.g. see \citealp[][in this volume and]{Vanni23} \citealp{Iwamoto2005a}
Because of their low explosion energy, indeed, these SNe are only able to expel a small fraction of Fe-peak elements from the innermost layers, yielding large amounts of carbon (and other light elements) with respect to iron.

Since stars form out of gas, moving to higher redshifts, we also expect to find very metal-poor gaseous environments primarily enriched by the first stars, thus showing a similar excess of carbon. To this aim, the chemical traces of Pop III stars have been searched for in quasar absorption lines probing dense neutral gas with [Fe/H]$< -2$, i.e. in very metal-poor Damped Lyman-$\alpha$ systems (DLAs) with neutral hydrogen column density $\log (N_{\rm HI}/ {\rm cm}^{-2})>20.3$. The initial claims of such a system \citep{Cooke2011} have since been corrected, and therefore no C-enhanced DLAs have been discovered so far \citep[][Pettini private comm.]{Dutta2014,Carswell2012}. 

Another approach is to investigate whether the chemical signature of the first stars can be identified in more diffuse high-z absorption systems, where star formation was likely interrupted early on, thus preventing contamination by subsequent generations of normal stars. To this end, we exploited the XQ-100 quasar legacy survey \citep{Lopez2016} to identify sub-DLAs and Lyman Limit Systems (LLSs) at redshift $z\approx 3-4$, i.e. absorption systems with H~I column densities in the range $17.2\leq \log(N_{\rm HI}/{\rm cm}^{-2})\leq20.3$. In \cite{Saccardi23}, we derived iron, carbon and other chemical abundances for these systems.

 Here we will compare our findings for gaseous systems with both observations of very metal-poor stars in the Local Group and theoretical models for Pop~III star enrichment. We refer the reader to \cite{Saccardi23} for an in-depth description of the different chemical abundances measured in high-z absorbers; and to \citealp[][this volume and submitted]{Vanni23} for a careful description of the theoretical model.

\section{CEMP-no stars in the Local Group}
\label{stars}
In the MW halo and nearby ultra-faint dwarf galaxies (UFDs), there are many known CEMP-no stars ($\rm [C/Fe]>+0.7$, $\rm[Ba/Fe]<0$), and a few have been also discovered in dwarf spheroidal (dSph) galaxies \citep[e.g.][]{Skuladottir2015}.
However, not all C-enhanced stars are CEMP-no, another class exists: the so called CEMP-s(/r) stars, which also exhibit an overabundance of neutron-capture elements ($\rm[Ba/Fe]>+1$). These CEMP-s stars are almost exclusively found in binary systems \citep[e.g.][]{Arentsen2019}, and their surplus of barium and carbon are expected to be acquired via an Asymptotic Giant Branch (AGB) companion during their lifetime \citep[e.g.][]{Abate2015}. Conversely, the C-excess of CEMP-no stars is expected to be a genuine representation of the environment of formation \citep[e.g.][]{Aguado2022}, which was likely polluted by the first stars \citep[e.g. see][in this volume]{Vanni23}. 

Ultimately, we need to find CEMP-no stars to indirectly study the first stellar generations. But how can we identify these stars without measuring barium? \cite{Spite2013} first demonstrated that we can discriminate among CEMP-no and CEMP-s stars on the basis of their absolute carbon abundance\footnote{$\tt A(C)=\log(N_{C}/N_{H})+12$}, A(C). Indeed, these two different populations dwell in two well defined regions: CEMP-s stars in the high-C band, A(C)$>7.4$, CEMP-no stars in the low-C band, A(C)$<7.4$ \citep[see also][]{Bonifacio2015a,Yoon2016}. 

\begin{figure}
\resizebox{\hsize}{!}{\includegraphics[clip=true]{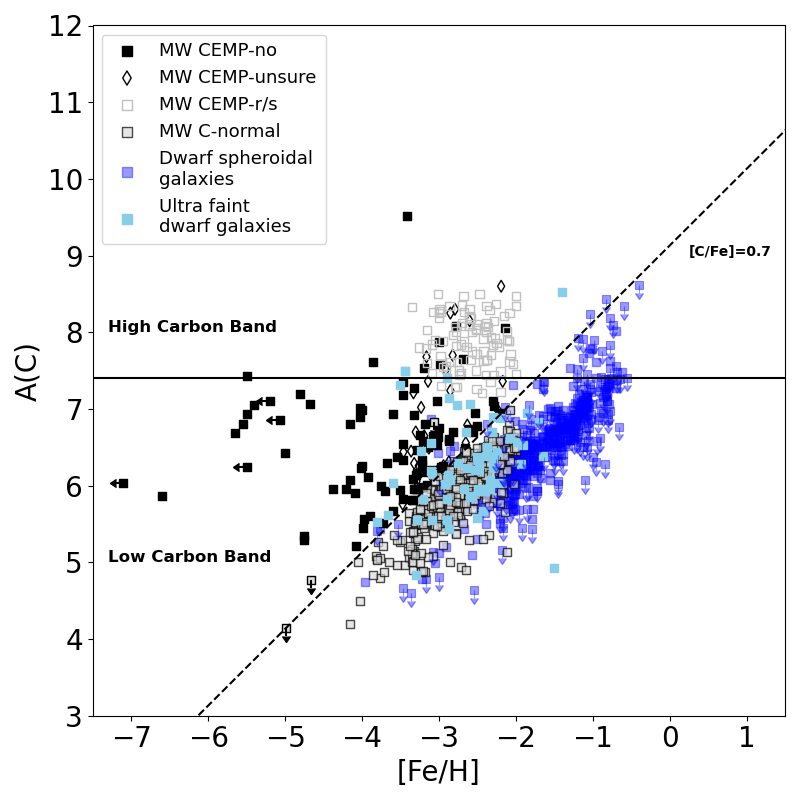}}
    \caption{\footnotesize Absolute carbon abundance, A(C), as a function of [Fe/H] for stars in UFDs (light blue squares), dSph galaxies (blue squares) and the MW (grey/black squares). The MW CEMP-no halo stars are shown as filled black symbols, C-normal halo stars as grey symbols, and CEMP-s(/r) stars as open symbols. Typical errors are $\sim$0.2\,dex.
    The horizontal line separates the low- and high-C bands, and the dashed line indicates $\rm[C/Fe]=+0.7$.}
    \label{fig:stars}
\end{figure}

In Fig.~\ref{fig:stars} we report the measured A(C) with [Fe/H] for stars in the Milky Way halo, UFDs, and dSph galaxies. The Figure contains data collected by \cite{Salvadori2015}, including new data in UFDs \citep{Ji2016,Spite2018} and newly discovered extremely metal-poor stars in the Milky Way halo \citep{Stankenburg2018,Francois2018,Bonifacio2018,Aguado2019,Gonzalez2020}. All carbon values are corrected to account for internal mixing processes \citep{Placco2014}.

It is clear from Fig.~\ref{fig:stars} that the division between the low-C and high-C bands is working remarkably well. In particular the low-C band is almost entirely populated by CEMP-no stars. 
Moreover, we note that CEMP-no stars become more dominant towards lower [Fe/H]. This is one of the reasons why they are believed to form in gaseous environment imprinted by the first stellar generations.

\section{CEMP-no absorbers at high-z}
\label{LLSs}
To search for the analogues of CEMP-no stars in high-z gaseous absorbers, we exploited the XQ-100 quasar legacy survey \citep{Lopez2016}, and selected a sample of 30 LLS and sub-DLAs at redshift $z\approx 3-4$, that we studied in detail \citep{Saccardi23}. We performed Voigt profile fitting of metal absorption features and Lyman lines in the quasar spectra to measure column densities. To derive relative abundances of different elements, we applied photoionization-model corrections to the measured ionic column densities by exploiting the open-source Cloudy \citep{Ferland2017}.

\begin{figure}
\resizebox{\hsize}{!}{\includegraphics[clip=true]{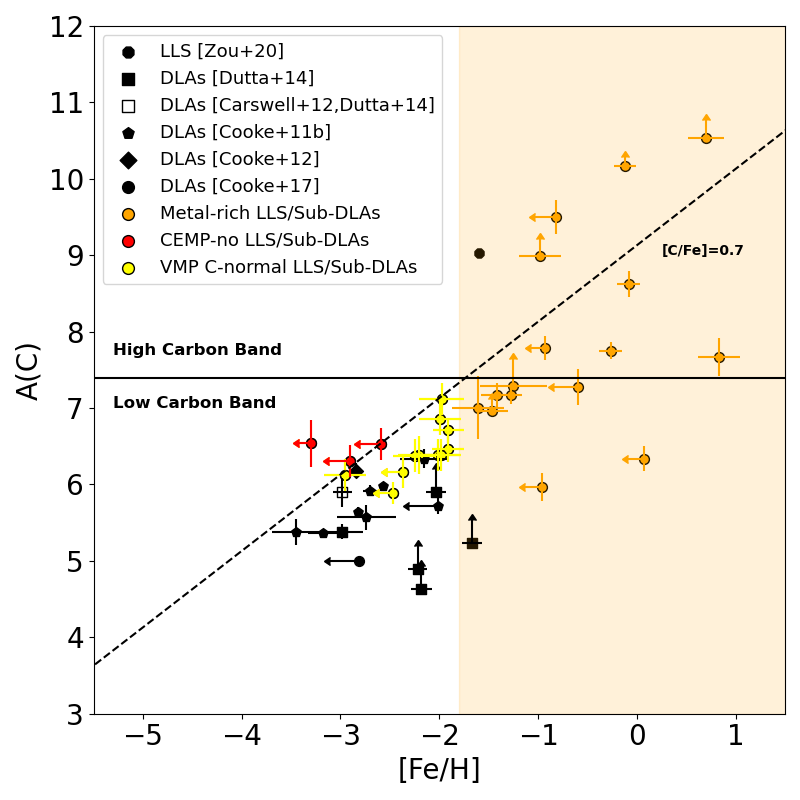}}
    \caption{\footnotesize Absolute carbon abundance as a function of [Fe/H] for the 30 sub-DLAs and LLSs of our sample (coloured circles) and for DLAs/LLSs from literature (black points, see labels and text for ref.).}
    \label{fig:LLS}
\end{figure}

Among these 30 absorption systems, we identified 14 very metal-poor absorbers, $\rm[Fe/H]<-2$, thereof three C-enhanced, $\rm[C/Fe]>+0.7$. 
Furthermore, we identified four C-enhanced systems at higher iron abundances, see Fig.~\ref{fig:LLS}.
We verified that the very metal-poor systems are also characterised by a low total metallicity ($Z<10^{-2}Z_{\odot}$), and that they are not simply iron-poor because of dust depletion \citep{Saccardi23}. Then, we concentrate on the three C-enhanced very metal-poor absorbers for which we can reasonably neglect the dust depletion \citep[e.g.][]{Vladilo1998,Vladilo2018,DeCia2018}.

To investigate the nature of our three bonafide CEMP systems, we show in Fig.~\ref{fig:LLS} the absolute carbon abundances with respect to [Fe/H], which is used as a key diagnostic tool to identify CEMP-no stars (Sec.~\ref{stars}). It is evident from Fig.~\ref{fig:LLS} that our three CEMP absorbers (red points) dwell in the low-C band, i.e. in the region where CEMP-no stars are typically found (see Fig.~\ref{fig:stars}). Note that we have only obtained $1-\sigma$ upper limits for [Fe/H] in these systems. However, a lower iron abundance will only increase their [C/Fe] abundance ratio.

\begin{figure}[]
\resizebox{\hsize}{!}{\includegraphics[clip=true]{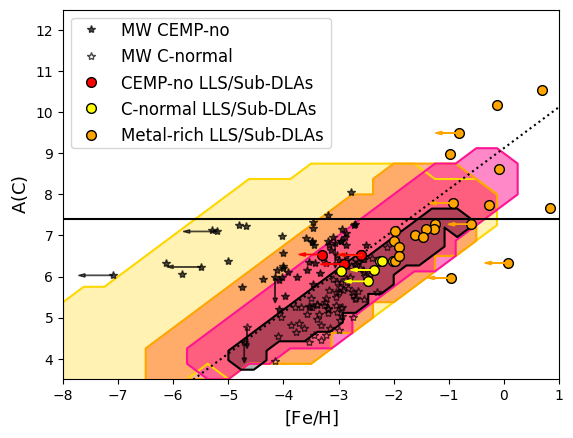}}
\caption{ \footnotesize Absolute carbon abundance as a function of [Fe/H] for an ISM imprinted by PopIII stars. Colours show different level of Pop~III enrichment: 100\% (yellow), 90\% (orange), 50\% (magenta), 10\% (gray). The dotted line shows $\rm[C/Fe]=+0.7$, the solid one separates the low-C from the high-C band.
}
\label{fig:model}
\end{figure}

In Fig.~\ref{fig:LLS} we also see that the C-normal very metal-poor absorbers (yellow points) reside where C-normal very metal-poor stars dwell. In this region, but at lower A(C) values, we can also see the very metal-poor DLAs from the literature \citep{Cooke2011,Cooke2012,Carswell2012,Dutta2014,Cooke2017}. Finally at $\rm[Fe/H]>-2$, we note that there are four C-enhanced absorbers \cite[plus one from][]{Zou2020} dwelling in the high-C band where CEMP-s stars are typically observed. At these metallicities, however, the dust depletion is no longer negligible, preventing us from drawing any strong conclusions.

\section{Origin of CEMP-no environments}
\label{results}

To understand the origin of the newly identified CEMP-no absorbers, we compare the observational results of \citep[][, Sec.~\ref{LLSs}]{Saccardi23}) to theoretical predictions. We exploit the simple and general parametric study by \citet{Salvadori2019}, which is further developed by Vanni et al. (submitted, and presented in this volume \citealp{Vanni23}. The goal of the model is to chemically characterize an interstellar medium (ISM) mainly imprinted by Pop~III stars, i.e. where the metals from Pop~III SNe account for $\geq 50\%$ of the total.

The model encodes the main unknowns related to early cosmic star formation and metal enrichment into three free parameters: the mass of gas converted into stars, or the star formation efficiency; metals retained in the ISM, parameterized with the dilution factor; and the mass fraction of PopIII metals with respect to the total, $\rm f_{PopIII}$. In \citealp{Vanni23} we demonstrated that either primordial faint SNe or core-collapse SNe are required to explain CEMP-no stars. Hence here we only focus on this class of low-energy primordial SNe.

In Fig. \ref{fig:model} we compare the predicted A(C) and [Fe/H] values for an ISM imprinted by low-energy ($\rm E\leq10^{51}$~erg) Pop~III supernovae with the chemical abundances measured in the gas of our high-z sub-DLAs and LLSs and in the stars of the MW halo. The colours denote different fraction of metals from Pop~III SNe: from the yellow area, where the ISM is completely imprinted by them ($\rm f_{popIII}=100\%$), to the gray area, where normal Pop~II stars exploding as core-collapse SNe dominate the ISM enrichment ($\rm f_{popIII}=10\%$). 

We can see in Fig.~\ref{fig:LLS} that our CEMP-no absorbers, which have moderately high [C/Fe] values, are consistent with a $\geq 50\%$ enrichment from Pop~III stars and thus can be considered gaseous fossils from the early Universe. 

\section{Conclusions}
\label{conclu}
We have shown that the newly discovered C-enhanced very metal-poor absorption systems most likely hold the chemical signature of the first stars that exploded as low-energy SNe ($\rm E\leq10^{51}$~erg), and provided $\geq50\%$ of their metals. These systems are thus the gaseous $z\approx 3-4$ analogues of present-day CEMP-no stars. This is further supported when their full abundance pattern is taken into account \citep{Saccardi23}.
 
The discovery of these gaseous CEMP-no systems suggests that optically thick, relatively diffuse absorbers, can preserve the chemical signatures of the first stellar generations. These absorbers are likely too diffuse to sustain star formation, a key requirement to prevent further chemical pollution and safegard the signature of the first stars \citep{Salvadori2012}. 

Presently we are limited by the relatively low resolution and signal-to-noise ratio of available observations. However, with the advent of ANDES for ELT \citep{Marconi2022} we shall be able to carry out more detailed studies of these absorbers at higher precision, detecting key elements like Zn \citep{Salvadori2019}. By exploiting the Pop~III star traces left in both {\it far-away gas} and in {\it nearby stars}, we will thus manage to understand the nature of the first stars and galaxies.  

\begin{acknowledgements}
This project received funding from the European Research Council (ERC) under the European Union’s Horizon 2020 research and innovation programme (grant agreement No 804240, PI S. Salvadori). S.S., V.D., and I.V. acknowledge support from the PRIN-MIUR17, prot. n. 2017T4ARJ5.
\end{acknowledgements}

\bibliographystyle{aa}
\bibliography{Main}

\begin{thebibliography}{40}
\expandafter\ifx\csname natexlab\endcsname\relax\def\natexlab#1{#1}\fi

\bibitem[{{Abate} {et~al.}(2015){Abate}, {Pols}, {Izzard}, \&
  {Karakas}}]{Abate2015}
{Abate}, C., {Pols}, O.~R., {Izzard}, R.~G., \& {Karakas}, A.~I. 2015, \aap,
  581, A22

\bibitem[{{Aguado} {et~al.}(2022){Aguado}, {Molaro}, {Caffau}, {Gonz{\'a}lez
  Hern{\'a}ndez}, {Zapatero Osorio}, {Bonifacio}, {Allende Prieto}, {Rebolo},
  {Damasso}, {Su{\'a}rez Mascare{\~n}o}, {Howell}, {Furlan}, {Cristiani},
  {Cupani}, {Di Marcantonio}, {D'Odorico}, {Lovis}, {Martins}, {Milakovic},
  {Murphy}, {Nunes}, {Pepe}, {Santos}, {Schmidt}, \& {Sozzetti}}]{Aguado2022}
{Aguado}, D., {Molaro}, P., {Caffau}, E., {et~al.} 2022, arXiv e-prints,
  arXiv:2210.04910

\bibitem[{{Aguado} {et~al.}(2019){Aguado}, {Gonz{\'a}lez Hern{\'a}ndez},
  {Allende Prieto}, \& {Rebolo}}]{Aguado2019}
{Aguado}, D.~S., {Gonz{\'a}lez Hern{\'a}ndez}, J.~I., {Allende Prieto}, C., \&
  {Rebolo}, R. 2019, \apjl, 874, L21

\bibitem[{{Arentsen} {et~al.}(2019){Arentsen}, {Starkenburg}, {Shetrone},
  {Venn}, {Depagne}, \& {McConnachie}}]{Arentsen2019}
{Arentsen}, A., {Starkenburg}, E., {Shetrone}, M.~D., {et~al.} 2019, \aap, 621,
  A108

\bibitem[{{Beers} \& {Christlieb}(2005)}]{Beers2005}
{Beers}, T.~C. \& {Christlieb}, N. 2005, \araa, 43, 531

\bibitem[{{Bonifacio} {et~al.}(2015){Bonifacio}, {Caffau}, {Spite}, {Limongi},
  {Chieffi}, {Klessen}, {Fran{\c{c}}ois}, {Molaro}, {Ludwig}, {Zaggia},
  {Spite}, {Plez}, {Cayrel}, {Christlieb}, {Clark}, {Glover}, {Hammer}, {Koch},
  {Monaco}, {Sbordone}, \& {Steffen}}]{Bonifacio2015a}
{Bonifacio}, P., {Caffau}, E., {Spite}, M., {et~al.} 2015, \aap, 579, A28

\bibitem[{{Bonifacio} {et~al.}(2018){Bonifacio}, {Caffau}, {Spite}, {Spite},
  {Sbordone}, {Monaco}, {Fran{\c{c}}ois}, {Plez}, {Molaro}, {Gallagher},
  {Cayrel}, {Christlieb}, {Klessen}, {Koch}, {Ludwig}, {Steffen}, {Zaggia}, \&
  {Abate}}]{Bonifacio2018}
{Bonifacio}, P., {Caffau}, E., {Spite}, M., {et~al.} 2018, \aap, 612, A65

\bibitem[{{Bromm} {et~al.}(2001){Bromm}, {Ferrara}, {Coppi}, \&
  {Larson}}]{Bromm2001}
{Bromm}, V., {Ferrara}, A., {Coppi}, P.~S., \& {Larson}, R.~B. 2001, \mnras,
  328, 969

\bibitem[{{Carswell} {et~al.}(2012){Carswell}, {Becker}, {Jorgenson}, {Murphy},
  \& {Wolfe}}]{Carswell2012}
{Carswell}, R.~F., {Becker}, G.~D., {Jorgenson}, R.~A., {Murphy}, M.~T., \&
  {Wolfe}, A.~M. 2012, \mnras, 422, 1700

\bibitem[{{Cooke} {et~al.}(2012){Cooke}, {Pettini}, \& {Murphy}}]{Cooke2012}
{Cooke}, R., {Pettini}, M., \& {Murphy}, M.~T. 2012, \mnras, 425, 347

\bibitem[{{Cooke} {et~al.}(2011){Cooke}, {Pettini}, {Steidel}, {Rudie}, \&
  {Nissen}}]{Cooke2011}
{Cooke}, R., {Pettini}, M., {Steidel}, C.~C., {Rudie}, G.~C., \& {Nissen},
  P.~E. 2011, \mnras, 417, 1534

\bibitem[{{Cooke} {et~al.}(2017){Cooke}, {Pettini}, \& {Steidel}}]{Cooke2017}
{Cooke}, R.~J., {Pettini}, M., \& {Steidel}, C.~C. 2017, \mnras, 467, 802

\bibitem[{{De Cia} {et~al.}(2018){De Cia}, {Ledoux}, {Petitjean}, \&
  {Savaglio}}]{DeCia2018}
{De Cia}, A., {Ledoux}, C., {Petitjean}, P., \& {Savaglio}, S. 2018, \aap, 611,
  A76

\bibitem[{{Dutta} {et~al.}(2014){Dutta}, {Srianand}, {Rahmani}, {Petitjean},
  {Noterdaeme}, \& {Ledoux}}]{Dutta2014}
{Dutta}, R., {Srianand}, R., {Rahmani}, H., {et~al.} 2014, \mnras, 440, 307

\bibitem[{{Ferland} {et~al.}(2017){Ferland}, {Chatzikos}, {Guzm{\'a}n},
  {Lykins}, {van Hoof}, {Williams}, {Abel}, {Badnell}, {Keenan}, {Porter}, \&
  {Stancil}}]{Ferland2017}
{Ferland}, G.~J., {Chatzikos}, M., {Guzm{\'a}n}, F., {et~al.} 2017, \aa, 53,
  385

\bibitem[{{Fran{\c{c}}ois} {et~al.}(2018){Fran{\c{c}}ois}, {Caffau}, {Wanajo},
  {Aguado}, {Spite}, {Aoki}, {Aoki}, {Bonifacio}, {Gallagher}, {Salvadori}, \&
  {Spite}}]{Francois2018}
{Fran{\c{c}}ois}, P., {Caffau}, E., {Wanajo}, S., {et~al.} 2018, \aap, 619, A10

\bibitem[{{Gonz{\'a}lez Hern{\'a}ndez} {et~al.}(2020){Gonz{\'a}lez
  Hern{\'a}ndez}, {Aguado}, {Allende Prieto}, {Burgasser}, \&
  {Rebolo}}]{Gonzalez2020}
{Gonz{\'a}lez Hern{\'a}ndez}, J.~I., {Aguado}, D.~S., {Allende Prieto}, C.,
  {Burgasser}, A.~J., \& {Rebolo}, R. 2020, \apjl, 889, L13

\bibitem[{{Heger} \& {Woosley}(2010)}]{Heger2010}
{Heger}, A. \& {Woosley}, S.~E. 2010, \apj, 724, 341

\bibitem[{{Hirano} {et~al.}(2014){Hirano}, {Hosokawa}, {Yoshida}, {Umeda},
  {Omukai}, {Chiaki}, \& {Yorke}}]{Hirano2014}
{Hirano}, S., {Hosokawa}, T., {Yoshida}, N., {et~al.} 2014, \apj, 781, 60

\bibitem[{{Iwamoto} {et~al.}(2005){Iwamoto}, {Umeda}, {Tominaga}, {Nomoto}, \&
  {Maeda}}]{Iwamoto2005a}
{Iwamoto}, N., {Umeda}, H., {Tominaga}, N., {Nomoto}, K., \& {Maeda}, K. 2005,
  Science, 309, 451

\bibitem[{{Ji} {et~al.}(2016){Ji}, {Frebel}, {Ezzeddine}, \& {Casey}}]{Ji2016}
{Ji}, A.~P., {Frebel}, A., {Ezzeddine}, R., \& {Casey}, A.~R. 2016, \apjl, 832,
  L3

\bibitem[{{L{\'o}pez} {et~al.}(2016){L{\'o}pez}, {D'Odorico}, {Ellison},
  {Becker}, {Christensen}, {Cupani}, {Denney}, {P{\^a}ris}, {Worseck}, {Berg},
  {Cristiani}, {Dessauges-Zavadsky}, {Haehnelt}, {Hamann}, {Hennawi},
  {Ir{\v{s}}i{\v{c}}}, {Kim}, {L{\'o}pez}, {Lund Saust}, {M{\'e}nard},
  {Perrotta}, {Prochaska}, {S{\'a}nchez-Ram{\'\i}rez}, {Vestergaard}, {Viel},
  \& {Wisotzki}}]{Lopez2016}
{L{\'o}pez}, S., {D'Odorico}, V., {Ellison}, S.~L., {et~al.} 2016, \aap, 594,
  A91

\bibitem[{{Marconi} {et~al.}(2022){Marconi}, {Abreu}, {Adibekyan}, {Alberti},
  {Albrecht}, {Alcaniz}, {Aliverti}, {Allende Prieto}, {Alvarado G{\'o}mez},
  {Amado}, {Amate}, {Andersen}, {Artigau}, {Baker}, {Baldini}, {Balestra},
  {Barnes}, {Baron}, {Barros}, {Bauer}, {Beaulieu}, {Bellido-Tirado},
  {Benneke}, {Bensby}, {Bergin}, {Biazzo}, {Bik}, {Birkby}, {Blind}, {Boisse},
  {Bolmont}, {Bonaglia}, {Bonfils}, {Borsa}, {Brandeker}, {Brandner}, {Broeg},
  {Brogi}, {Brousseau}, {Brucalassi}, {Brynnel}, {Buchhave}, {Buscher},
  {Cabral}, {Calderone}, {Calvo-Ortega}, {Canto Martins}, {Cantalloube},
  {Carbonaro}, {Chauvin}, {Chazelas}, {Cheffot}, {Cheng}, {Chiavassa},
  {Christensen}, {Cirami}, {Cook}, {Cooke}, {Coretti}, {Covino}, {Cowan},
  {Cresci}, {Cristiani}, {Cunha Parro}, {Cupani}, {D'Odorico}, {de Castro
  Le{\~a}o}, {De Cia}, {De Medeiros}, {Debras}, {Debus}, {Demangeon},
  {Dessauges-Zavadsky}, {Di Marcantonio}, {Dionies}, {Doyon}, {Dunn},
  {Ehrenreich}, {Faria}, {Feruglio}, {Fisher}, {Fontana}, {Fumagalli}, {Fusco},
  {Fynbo}, {Gabella}, {Gaessler}, {Gallo}, {Gao}, {Genolet}, {Genoni},
  {Giacobbe}, {Giro}, {Gon{\c{c}}alves}, {Gonzalez}, {Gonz{\'a}lez
  Hern{\'a}ndez}, {Gracia T{\'e}mich}, {Haehnelt}, {Haniff}, {Hatzes},
  {Helled}, {Hoeijmakers}, {Huke}, {J{\"a}rvinen}, {J{\"a}rvinen}, {Kaminski},
  {Korn}, {Kouach}, {Kowzan}, {Kreidberg}, {Landoni}, {Lanotte}, {Lavail},
  {Li}, {Liske}, {Lovis}, {Lucatello}, {Lunney}, {MacIntosh}, {Madhusudhan},
  {Magrini}, {Maiolino}, {Malo}, {Man}, {Marquart}, {Marques}, {Martins},
  {Martins}, {Maslowski}, {Mason}, {Mason}, {McCracken}, {Mergo}, {Micela},
  {Mitchell}, {Molli{\`e}re}, {Monteiro}, {Montgomery}, {Mordasini}, {Morin},
  {Mucciarelli}, {Murphy}, {N'Diaye}, {Neichel}, {Niedzielski}, {Niemczura},
  {Nortmann}, {Noterdaeme}, {Nunes}, {Oggioni}, {Oliva}, {{\"O}nel}, {Origlia},
  {{\"O}stlin}, {Palle}, {Papaderos}, {Pariani}, {Pe{\~n}ate Castro}, {Pepe},
  {Perreault Levasseur}, {Petit}, {Pino}, {Piqueras}, {Pollo}, {Poppenhaeger},
  {Quirrenbach}, {Rauscher}, {Rebolo}, {Redaelli}, {Reffert}, {Reid},
  {Reiners}, {Richter}, {Riva}, {Rivoire}, {Rodr{\'\i}guez-L{\'o}pez},
  {Roederer}, {Romano}, {Rousseau}, {Rowe}, {Salvadori}, {Santos}, {Santos
  Diaz}, {Sanz-Forcada}, {Sarajlic}, {Sauvage}, {Sch{\"a}fer}, {Schiavon},
  {Schmidt}, {Selmi}, {Sivanandam}, {Sordet}, {Sordo}, {Sortino}, {Sosnowska},
  {Sousa}, {Stempels}, {Strassmeier}, {Su{\'a}rez Mascare{\~n}o}, {Sulich},
  {Sun}, {Tanvir}, {Tenegi-Sangin{\'e}s}, {Thibault}, {Thompson}, {Tozzi},
  {Turbet}, {Vall{\'e}e}, {Varas}, {Venn}, {V{\'e}ran}, {Verma}, {Viel},
  {Wade}, {Waring}, {Weber}, {Weder}, {Wehbe}, {Weingrill}, {Woche}, {Xompero},
  {Zackrisson}, {Zanutta}, {Zapatero Osorio}, {Zechmeister}, \&
  {Zimara}}]{Marconi2022}
{Marconi}, A., {Abreu}, M., {Adibekyan}, V., {et~al.} 2022, in SPIE Conference
  Series, Vol. 12184, Ground-based and Airborne Instrumentation for Astronomy
  IX, ed. C.~J. {Evans}, J.~J. {Bryant}, \& K.~{Motohara}, 1218424

\bibitem[{{Placco} {et~al.}(2014){Placco}, {Frebel}, {Beers}, \&
  {Stancliffe}}]{Placco2014}
{Placco}, V.~M., {Frebel}, A., {Beers}, T.~C., \& {Stancliffe}, R.~J. 2014,
  \apj, 797, 21

\bibitem[{{Rossi} {et~al.}(2021){Rossi}, {Salvadori}, \&
  {Sk{\'u}lad{\'o}ttir}}]{Rossi2021}
{Rossi}, M., {Salvadori}, S., \& {Sk{\'u}lad{\'o}ttir}, {\'A}. 2021, \mnras,
  503, 6026

\bibitem[{{Saccardi} {et~al.}(2023){Saccardi}, {Salvadori}, {D'Odorico},
  {Cupani}, {Fumagalli}, {Berg}, {Becker}, {Ellison}, \& {Lopez}}]{Saccardi23}
{Saccardi}, A., {Salvadori}, S., {D'Odorico}, V., {et~al.} 2023, \apj, 948, 35

\bibitem[{{Salvadori} {et~al.}(2019){Salvadori}, {Bonifacio}, {Caffau},
  {Korotin}, {Andreevsky}, {Spite}, \& {Sk{\'u}lad{\'o}ttir}}]{Salvadori2019}
{Salvadori}, S., {Bonifacio}, P., {Caffau}, E., {et~al.} 2019, \mnras, 487,
  4261

\bibitem[{{Salvadori} \& {Ferrara}(2012)}]{Salvadori2012}
{Salvadori}, S. \& {Ferrara}, A. 2012, \mnras, 421, L29

\bibitem[{{Salvadori} {et~al.}(2015){Salvadori}, {Sk{\'u}lad{\'o}ttir}, \&
  {Tolstoy}}]{Salvadori2015}
{Salvadori}, S., {Sk{\'u}lad{\'o}ttir}, {\'A}., \& {Tolstoy}, E. 2015, \mnras,
  454, 1320

\bibitem[{{Sk{\'u}lad{\'o}ttir} {et~al.}(2015){Sk{\'u}lad{\'o}ttir}, {Tolstoy},
  {Salvadori}, {Hill}, {Pettini}, {Shetrone}, \&
  {Starkenburg}}]{Skuladottir2015}
{Sk{\'u}lad{\'o}ttir}, {\'A}., {Tolstoy}, E., {Salvadori}, S., {et~al.} 2015,
  \aap, 574, A129

\bibitem[{{Spite} {et~al.}(2013){Spite}, {Caffau}, {Bonifacio}, {Spite},
  {Ludwig}, {Plez}, \& {Christlieb}}]{Spite2013}
{Spite}, M., {Caffau}, E., {Bonifacio}, P., {et~al.} 2013, \aap, 552, A107

\bibitem[{{Spite} {et~al.}(2018){Spite}, {Spite}, {Fran{\c{c}}ois},
  {Bonifacio}, {Caffau}, \& {Salvadori}}]{Spite2018}
{Spite}, M., {Spite}, F., {Fran{\c{c}}ois}, P., {et~al.} 2018, \aap, 617, A56

\bibitem[{{Starkenburg} {et~al.}(2018){Starkenburg}, {Aguado}, {Bonifacio},
  {Caffau}, {Jablonka}, {Lardo}, {Martin}, {S{\'a}nchez-Janssen}, {Sestito},
  {Venn}, {Youakim}, {Allende Prieto}, {Arentsen}, {Gentile}, {Gonz{\'a}lez
  Hern{\'a}ndez}, {Kielty}, {Koppelman}, {Longeard}, {Tolstoy}, {Carlberg},
  {C{\^o}t{\'e}}, {Fouesneau}, {Hill}, {McConnachie}, \&
  {Navarro}}]{Stankenburg2018}
{Starkenburg}, E., {Aguado}, D.~S., {Bonifacio}, P., {et~al.} 2018, \mnras,
  481, 3838

\bibitem[{{Susa} {et~al.}(2014){Susa}, {Hasegawa}, \& {Tominaga}}]{Susa2014}
{Susa}, H., {Hasegawa}, K., \& {Tominaga}, N. 2014, \apj, 792, 32

\bibitem[{{Tolstoy} {et~al.}(2009){Tolstoy}, {Hill}, \& {Tosi}}]{Tolstoy2009}
{Tolstoy}, E., {Hill}, V., \& {Tosi}, M. 2009, \araa, 47, 371

\bibitem[{{Vanni} {et~al.}(2023){Vanni}, {Salvadori}, \&
  {Sk{\'u}lad{\'o}ttir}}]{Vanni23}
{Vanni}, I., {Salvadori}, S., \& {Sk{\'u}lad{\'o}ttir}, {\'A}. 2023, arXiv
  e-prints, arXiv:2305.02358

\bibitem[{{Vladilo}(1998)}]{Vladilo1998}
{Vladilo}, G. 1998, \apj, 493, 583

\bibitem[{{Vladilo} {et~al.}(2018){Vladilo}, {Gioannini}, {Matteucci}, \&
  {Palla}}]{Vladilo2018}
{Vladilo}, G., {Gioannini}, L., {Matteucci}, F., \& {Palla}, M. 2018, \apj,
  868, 127

\bibitem[{{Yoon} {et~al.}(2016){Yoon}, {Beers}, {Placco}, {Rasmussen},
  {Carollo}, {He}, {Hansen}, {Roederer}, \& {Zeanah}}]{Yoon2016}
{Yoon}, J., {Beers}, T.~C., {Placco}, V.~M., {et~al.} 2016, \apj, 833, 20

\bibitem[{{Zou} {et~al.}(2020){Zou}, {Petitjean}, {Noterdaeme}, {Ledoux},
  {Srianand}, {Jiang}, \& {Krogager}}]{Zou2020}
{Zou}, S., {Petitjean}, P., {Noterdaeme}, P., {et~al.} 2020, \apj, 901, 105

\end{thebibliography}

\end{document}